\documentstyle[11pt,newpasp,twoside]{article}
\def\edcomment#1{\iffalse\marginpar{\raggedright\sl#1\/}\else\relax\fi}
\reversemarginpar

\def\beq{\begin{equation}}
\def\eeq{\end{equation}}
\def\Fig#1{Fig.~\ref{fig:#1}}
\def\equ#1{eq.~\ref{eq:#1}}

\def\etal{{\it et al.\ }}
\def\ltsima{$\; \buildrel < \over \sim \;$}
\def\lsim{\lower.5ex\hbox{\ltsima}}
\def\gtsima{$\; \buildrel > \over \sim \;$}
\def\gsim{\lower.5ex\hbox{\gtsima}}
\def\dd{{d}}

\def\Rv{R_{\rm vir}}

\def\rc{r_{\rm c}}
\def\ellc{\ell_{\rm c}}
\def\sigc{\sigma_{\rm c}}
\def\rhoc{\rho_{\rm c}}
\def\mnu{\nu}
\def\aas{\alpha_{\rm a}}

\def\pmb#1{\setbox0=\hbox{#1}%
\kern-.025em\copy0\kern-\wd0
\kern.05em\copy0\kern-\wd0
\kern-.025em\raise.0433em\box0}
\def\bell{\pmb{$\ell$}}

\begin{document}
\title{Galactic Halo Cusp versus Core: Tidal Effects in Mergers}
\author{Avishai Dekel, Jonathan Devor \& Itai Arad}
\affil{Racah Institute of Physics, The Hebrew University of Jerusalem}

\begin{abstract}
We show how the buildup of halos by merging satellites forces 
an inner cusp, with a density profile $\rho \propto r^{-\alpha}$
where $\alpha \rightarrow \aas \gsim 1$. 
Our analysis is based on a new prescription for tidal stripping  
as a function of $\alpha(r)$, using a simple toy model which matches N-body
simulations.
In a core of $\alpha < 1$ there is tidal compression rather than stripping and
the satellites sink towards the halo center, causing a rapid steepening of
the profile to $\alpha > 1$.  Where $\alpha > 1$,  
the stripping of each satellite shell is preceded by gradual puffing up,
which makes the stripping more efficient at larger $\alpha$, causing
flattening where $\alpha$ is large enough. Therefore, we can show using linear 
perturbation analysis that a sequence of mergers slowly leads to a fixed point 
$\alpha(r)=\aas$.  This result implies that a cusp is enforced as long as 
enough satellite material makes it into the inner halo and is deposited there. 
We conclude that in order to maintain a flat core, as indicated by observations,
satellites must be disrupted outside the core,
e.g., because of puffing up due to baryonic feedback effects. 
\end{abstract}

\section{Introduction}
\label{sec:intro}

The `standard' model of cosmology, CDM, which assumes hierarchical buildup of
structure, is facing difficulties in explaining
observed properties of galaxies, such as the number density of dwarfs
(e.g., Klypin \etal 1999b),
the angular-momentum crisis
(e.g., Navarro \& Steinmetz 2000), and the cusp/core problem.
Our approach in addressing these problems within CDM is to isolate and model in 
{\it simple} physical terms the key relevant processes, as a guide for 
possible solutions. 
We first model the buildup of dark-matter (DM) halos in N-body simulations
based on tidal effects,
and then incorporate the inevitable
baryonic feedback processes in an attempt to explain 
the apparent discrepancies.
We address the angular momentum problem in Maller \& Dekel (2002) and
Dekel \& Maller (2002), and summarize here our progress in the
cusp/core problem (Dekel \& Devor 2002; Dekel \etal 2002).

Cosmological N-body simulations have revealed that the density profiles
of DM halos scatter about a universal shape,
$\rho (r) = \rhoc \, (r/\rc)^{-\alpha} \, (1+r/\rc)^{\alpha-3}$,
with an inner {\it cusp} of slope $-\alpha$.
Navarro, Frenk \& White (1995, NFW) found this function, with 
$\alpha \simeq 1$,
to be a good fit in the range $(0.01-1)\Rv$
for different hierarchical cosmological scenarios.
High-resolution simulations of a few individual halos
(Moore \etal 2001; Klypin \etal 2001)
found that the cusp could be as steep as $\alpha \simeq 1.5$,
though it may flatten towards $\alpha \simeq 1$ at $r<0.01\Rv$ 
(private comm. with Navarro, Frenk, Springel \& White).  
While the formation of a cusp with $1 \leq \alpha \leq 1.5$ has been
established in the simulations, a basic understanding of its 
origin is still lacking.
An even more intriguing puzzle is introduced by observations of low
surface-brightness galaxies,
whose centers are dominated by their DM halos, which indicate
flatter inner {\it cores} with $\alpha \simeq 0$ (de Block \etal 2001).
This seems to introduce a challenge to the CDM paradigm.

In the first two sections we develop a toy model for tidal stripping 
and test it against an N-body simulation.
In 
\S 2
we describe the compression and rapid steepening in a core.
In 
\S 3
we address stripping where $\alpha > 1$.
In 
\S 4
we analyze the convergence to an asymptotic profile.
In 
\S 5
we discuss our results.

\section{Core into Cusp due to Tidal Compression}
\label{sec:compress}

We consider a fixed spherical halo of mean density profile
$\bar\rho(r)\!\propto\!M(r)/r^3$.
Denote
$\alpha(r) \equiv - {d\ln \bar\rho / d\ln r}$
such that locally $\bar\rho \propto r^{-\alpha}$,
with $\alpha(r)$ constant or increasing in the range
$0 \leq \alpha \leq 3$.
Consider a satellite at $r$, spiraling into the halo under gravity and
dynamical friction.  The maximum tidal force by the halo on a
unit mass at satellite radius $\bell$ ($\ell \ll r$) is
\beq
{\bf F}_{\rm tide} = {\mu (r) \ell \over r^3} \hat{\bell} ,
\quad \mu (r) = {2M(r)-r\, {\dd M \over \dd r}}
=(\alpha-1)\,M(r) \, .
\label{eq:tide} 
\eeq
The maximum is obtained along the line connecting the halo centers, but
while the radial component is smaller in all other directions,
it is of the same sign.
The familiar case is of a point mass, $\alpha=3$,
for which the tidal force is pulling outwards from the satellite.
For flatter halo slopes it becomes weaker $\propto\!(\alpha-1)$.
An important feature that is often overlooked is that when the halo
density profile is flat enough, $\alpha<1$, the tidal force reverses
direction into compression, resulting in accretion rather than stripping.
Note that the critical slope of unity coincides with
the cusp of the NFW profile, and that in a core,
$\alpha \sim 0$, the tides induce strong compression.

This effect can be demonstrated using an N-body simulation of a merger
(following Mihos \& Hernquist 1996). We use a total of $10^5$ DM 
particles, the mass ratio is 1:10, the satellite spirals in on a quasi-circular
orbit, and the initial profile is a truncated isothermal sphere with a core,
where $\alpha$ ranges in practice from 0.6 to 2.8.
In 
Fig.~1
(left), the stripping of each satellite shell, marked by the 
onset of a steep rise, can be identified with a halo radius $r$, 
with $\alpha(r)$ decreasing from 3 downwards as one moves from outer 
shells inwards.
However, as the stripping point approaches $\alpha \sim 1$,
inwards to the 30\% satellite mass shell, the stripping stops
and some shrinking can be seen instead.

The compression at $\alpha<1$ implies that any part of the satellite
which makes it intact into the halo core would sink towards the center without
further stripping.  This should cause a rapid steepening of the profile to 
$\alpha > 1$, as seen in 
Fig.~1
(right).
The NFW inner slope of $\alpha =1$ is thus a robust lower bound, as we know 
from cosmological simulations; a flatter density core cannot survive as
long as satellites deposit enough mass in the inner halo.

\begin{figure} 
\vskip 4.5truecm
 \includegraphics{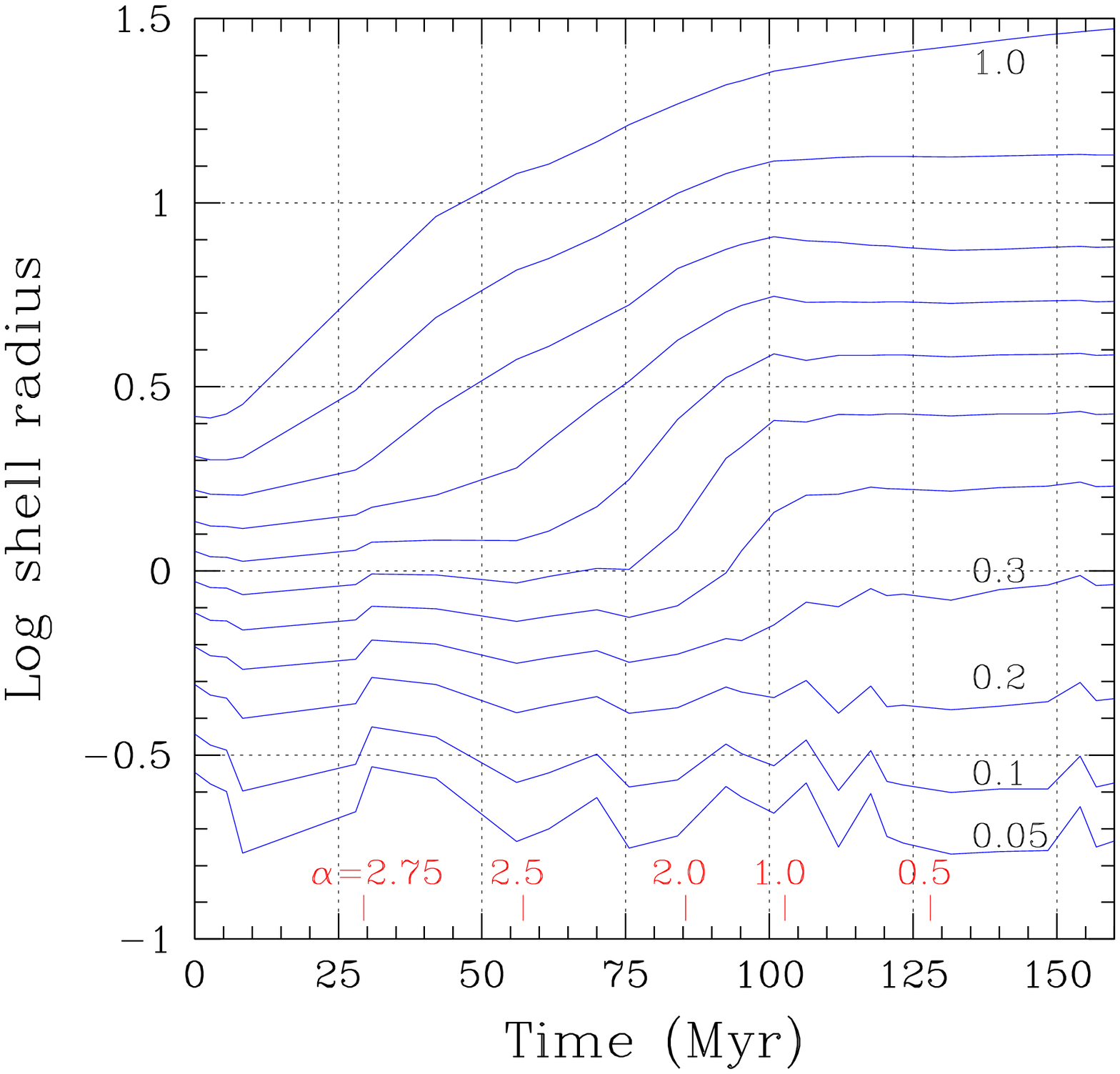}
 \includegraphics{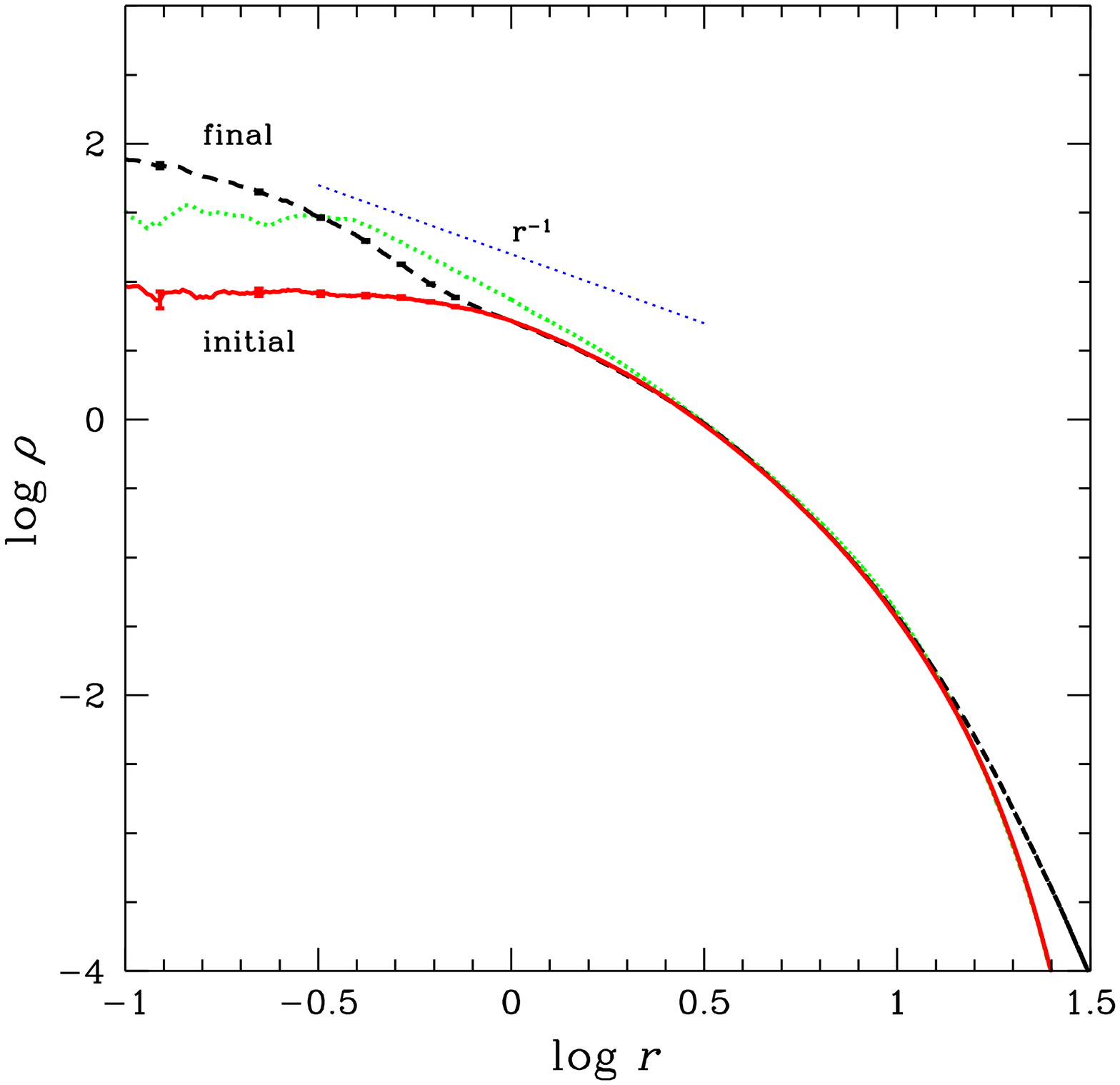}
\caption{
{\it Left}: Evolution of the satellite mass profile during the merger
simulation.
Shown are the radii of spheres about the satellite center,
encompassing given fractions of the satellite mass.
Marked along the x-axis are values of $\alpha(r)$ corresponding to the
momentary position of the satellite center in the halo.
The inner 25\% of the satellite mass is never torn away --- it enters
intact the $\alpha \leq 1$ core.
{\it Right}:
Halo density profile before (solid) and after (dashed) the merger.
The compact inner satellite sinks into the halo core and produces a cusp
with $\alpha \geq 1$.
}
\label{fig:cusp}
\end{figure}

\section{Tidal Puffing-Up and Stripping at $\alpha >1$}
\label{sec:puff}

At $\alpha>1$, the effects are more subtle.
Let $\ell$ mark shell radii within the unperturbed satellite,
whose mean-density profile is $\bar{\sigma}(\ell) \propto m(\ell)/\ell^3$.
Assume that when it is at halo radius $r$, mass is lost beyond a momentary
stripping radius $\ell(r)$ and is added to the halo at $r$ (on average).
We wish to determine the correspondence between $\ell$ and $r$ at stripping.
Define $\psi (r,\ell) \equiv \bar\rho(r) / \bar\sigma(\ell)$.
Traditionally, the stripping radius is assumed to be determined by the
resonance condition $\psi(r,\ell)=1$, but this ignores the earlier
effects of tides on the satellite structure.
A key new feature in our analysis is that, as $r$ decreases,
the tides stretch the satellite orbits, which can be modeled as
an effective puffing up of the relevant shells before they are being
torn away.  We define for every shell $\ell$ when the satellite is at $r$
a momentary {\it puffing factor} by
$p(r,\ell) \equiv \ell_{\rm p}/\ell$, where
$\ell_{\rm p}(r)$ is the momentary shell radius.
One can then show that the {\it resonance
condition}, for $r$ and $\ell$ at stripping, becomes
\beq
\psi=\alpha^{-1} p^{-3} \, .
\eeq
In the regime where $\alpha>1$, we expect puffing, $p>1$, so the
corrected resonance condition implies $\psi<1$, and for large $\alpha$
even $\psi\ll 1$. This means more efficient stripping
compared to the old condition ignoring puffing.
To obtain an explicit stripping condition we wish to express $\psi$ as a
function of $\alpha$, so we need to estimate how $p(\ell)$ evolves
as the satellite falls into smaller $r$ positions.
By applying an adiabatic invariant, we obtain a {\it puffing equation}
for any
shell $\ell$ when it is at $r$:
\beq
p - (\alpha-1) \psi\, p^4 =1 \, .
\eeq
In the two equations above we have omitted geometrical factors of order
unity, to be calibrated later using simulations (see Dekel \& Devor 2002).
When we combine the above equations we obtain a new {\it stripping
condition}:
\beq
{\bar\rho(r) \over \bar\sigma(\ell)}
= \psi[\alpha(r)]
= \cases{ \alpha^{-4}     & $1 < \alpha < \alpha_{\rm c} \sim 1.4$ \cr
           0.1/(\alpha-1) & $\alpha > \alpha_{\rm c}$     } \, .
\label{eq:strip_con} 
\eeq
Given the slope profile $\alpha(r)$, it relates every satellite shell $\ell$ to
the position $r$ where it should be stripped.
The function $\psi(\alpha)$ is thus predicted to decrease monotonically
as a function of $\alpha$, towards values of order 0.1-0.2 at
$\alpha \sim 2$ and below 0.1 at $\alpha=3$, 
Fig.~2
(left).
This implies that the stripping process is more efficient for steeper
halo profiles.
We also implement in the deposit prescription additional effects near
and below $\alpha =1$, due to the finite size of the satellite
and the tidal compression in the inner halo. These effects may
cause deposit of satellite material before the stripping condition
is fulfilled.

\begin{figure} 
\vskip 4.5 truecm
\includegraphics{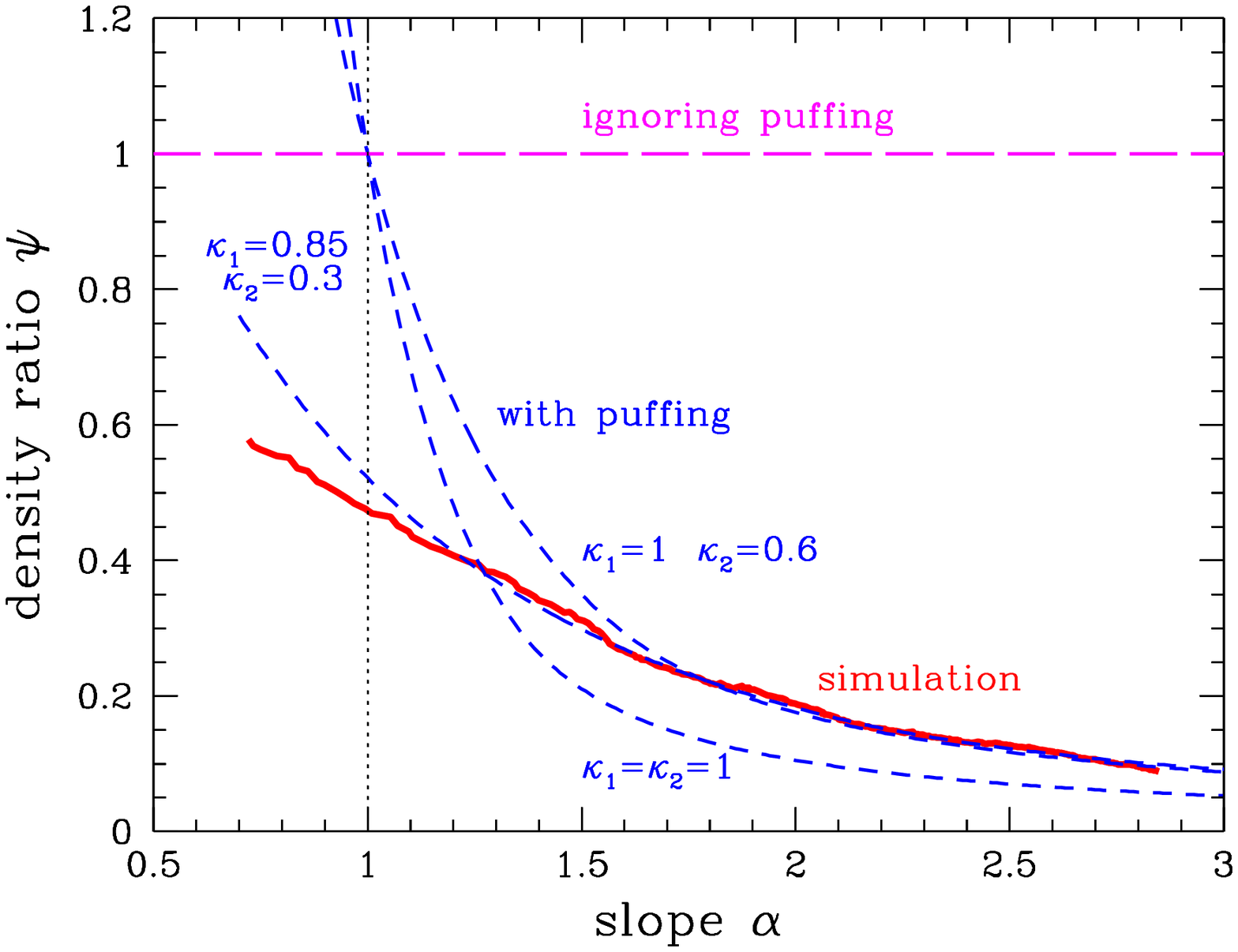}
\includegraphics{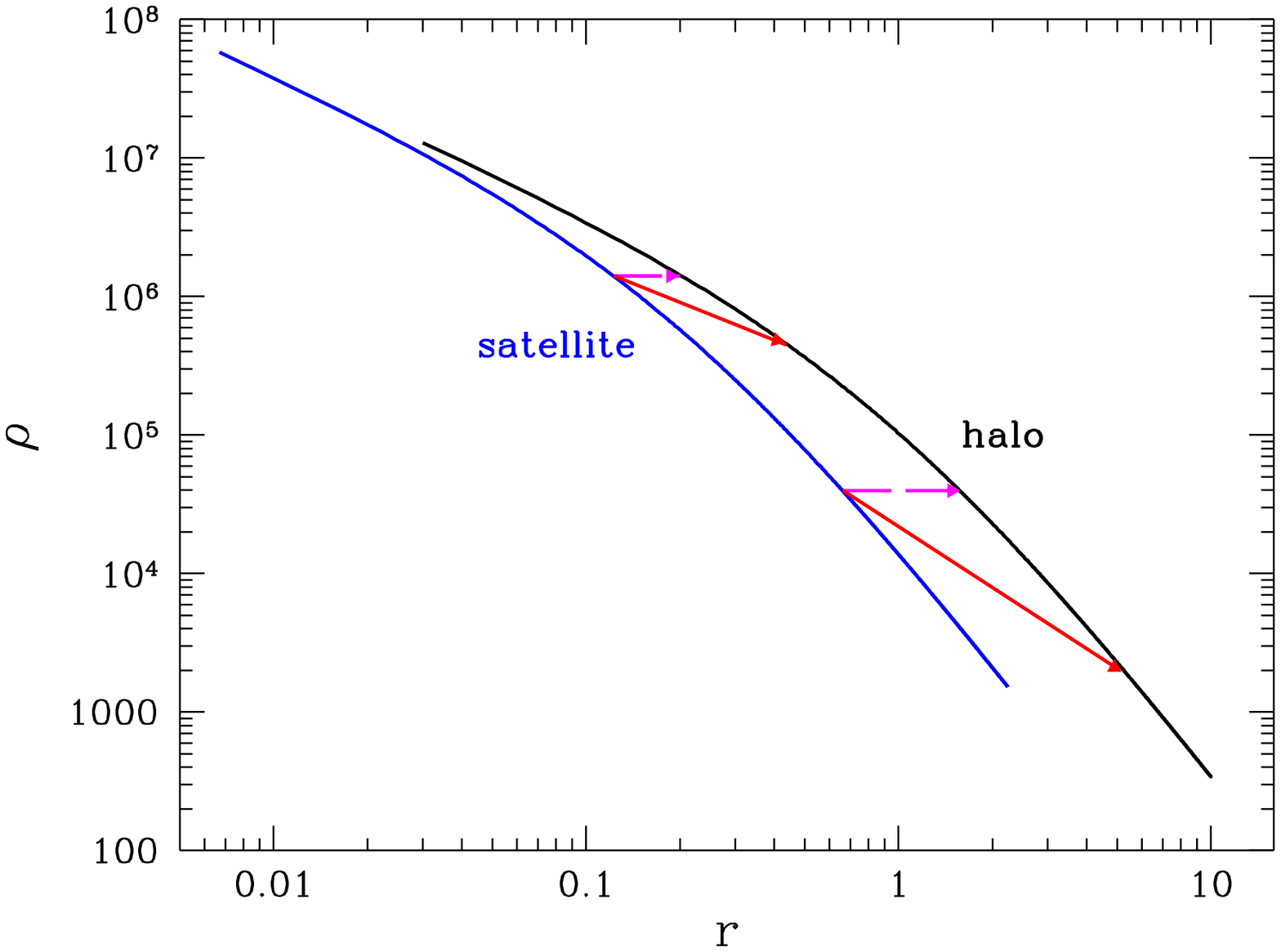}
\caption{       
{\it Left}: 
Density ratio $\psi$ at stripping as a function of halo slope $\alpha$.
The predictions of the model are shown (dashed)
for three different choices of the geometrical factors.
The no-puffing prediction is $\psi=1$.
The result of the N-body simulation (solid) demonstrates
that the puffing model can provide a good approximation.
{\it Right}:
A schematic illustration of satellite mass deposit in the halo.
Shown are an NFW halo profile $\bar\rho(r)$, and a homologous satellite
$\bar\sigma(\ell)$ properly shifted to the left and upwards.
The arrows connect shell radii $\ell$ to the halo radii where they are
deposited $r$.
The horizontal dashed arrows refer to stripping when puffing is ignored,
$\psi(\alpha)=1$. This would steepen the profile, as steep regions of
$\bar\sigma(\ell)$ are deposited at flatter regions of $\bar\rho(r)$.
The solid arrows illustrate realistic stripping after puffing.
The vertical displacements, which grow with $r$, refer to
$\psi(\alpha)<1$.
The slope at $\ell$ may be flatter than the slope at $r$ such that the
mass tends to be deposited at larger $r$ and the result is flattening
of the halo profile.
          } 
\label{fig:psial}
\end{figure}

The puffing before stripping can be seen in the merger simulation, 
e.g., for the outer and intermediate shells in 
Fig.~1
(left).
The magnitude of the puffing is 30-50\%, as expected,
corresponding to a factor of 2-3 in density.
A similar effect has been qualitatively noticed in
simulations before (e.g. Klypin \etal 1999a, Fig.~6).
In the merger simulation, we measure the deposit radius $r(\ell)$ 
from the final distribution of stripped satellite mass about 
the halo center [can be read from 
\Fig{cusp} 
Fig.~1
(left)].  
The corresponding values of $\psi(r,\ell)$
are plotted against $\alpha(r)$ in 
Fig.~2
(left).
The qualitative agreement between the simulation result
and the model predictions indicates that despite the
crude approximations made, our very simplified model mimics the main
features of the tidal stripping and deposit process.
When puffing is ignored, $\psi=1$, the model clearly fails.

\section{Halo Asymptotic Profile}
\label{sec:profile}

If the stripping is described by a condition similar
to \equ{strip_con}, with $\psi(\alpha)$ a decreasing function,
the profile evolves slowly towards an asymptotic
stable power law with $\aas \gsim 1$.
We assume that the halo and satellite are drawn from a cosmological
distribution; they are homologous,
with their characteristic radii and densities scaling like 
$\ellc/\rc \propto m^{(1+\mnu)/3}$
and
$\sigc/\rhoc \propto  m^{-\mnu}$,
where $\mnu \simeq 0.33$ for $\Lambda$CDM.
Fig.~2
(right) helps understanding the origin of an asymptotic slope.
We write
$\bar\rho_{\rm final}(r) = \bar\rho(r) +\bar\sigma(\ell) {\ell^3/ r^3}$,
and obtain for the change of $\alpha$ in a merger
\beq
\Delta \alpha(r) 
\propto  {\dd \over \dd r} 
\left( {\bar\sigma(\ell) \over \bar\rho(r)} {\ell^3 \over r^3} \right)
\, .
\label{eq:dalpha} 
\eeq
One can see that every power law is a self-similar solution,
$\Delta\alpha(r)=0$, but not necessarily a stable one.
When $\alpha$ is increasing with $r$, $\ell/r$ is decreasing with $r$.
Thus, when puffing is ignored, $\psi=$const., one has
continuous steepening, $\Delta\alpha(r)>0$.
With realistic puffing, $1/\psi$ is increasing with
$r$, which can produce a stable fixed point at a certain asymptotic
value $\aas$,
where $\Delta\alpha=0$ and the second derivative is negative.
A rigorous linear perturbation analysis determines the rate of convergence
to $\aas$ and yields an equation for its value for a sequence of mergers
with the same mass ratio:
$$
\Delta\alpha \propto \alpha(\alpha - 3){\psi'(\alpha)}/{\psi(\alpha)}
+ 3 \ln[(m/M)^{-\mnu} \psi(\alpha)]  = 0 \,.
\label{eq:aas}
$$
The solutions are typically in the range $1 < \aas \leq 1.5$.

In order to test the linear analysis, we perform
toy simulations of the profile buildup by mergers,
where we implement the stripping recipe of 
\S 3
(replacing the crude stripping recipe used in earlier
work, e.g., Syre \& White 1998).
For given halo and satellite profiles, we solve the stripping equation
for $\ell(r)$ and add the stripped satellite mass to the halo accordingly.
Among other numerical complications, we implement a smoothing
scheme to ensure that $\alpha(r)$ remains monotonic.
We then follow a sequence of cosmological mergers and
study the evolution towards an asymptotic slope.
Fig.~3
(left) shows the convergence of $\alpha$ at a fixed $r$ to the
asymptotic value.
Fig.~3
(right) shows how the profile evolves through momentary profiles
which are probably more relevant for comparison with real halos at different
times during their buildup process.
These figures are for certain given values of the geometrical factors, 
the mass ratio and $\mnu$.  In Dekel \etal (2002) we address 
a sequence of mergers with a cosmological distribution of mass ratios,
and the robustness to the cosmological model.

\begin{figure} 
\vskip 4.5truecm
{\includegraphics{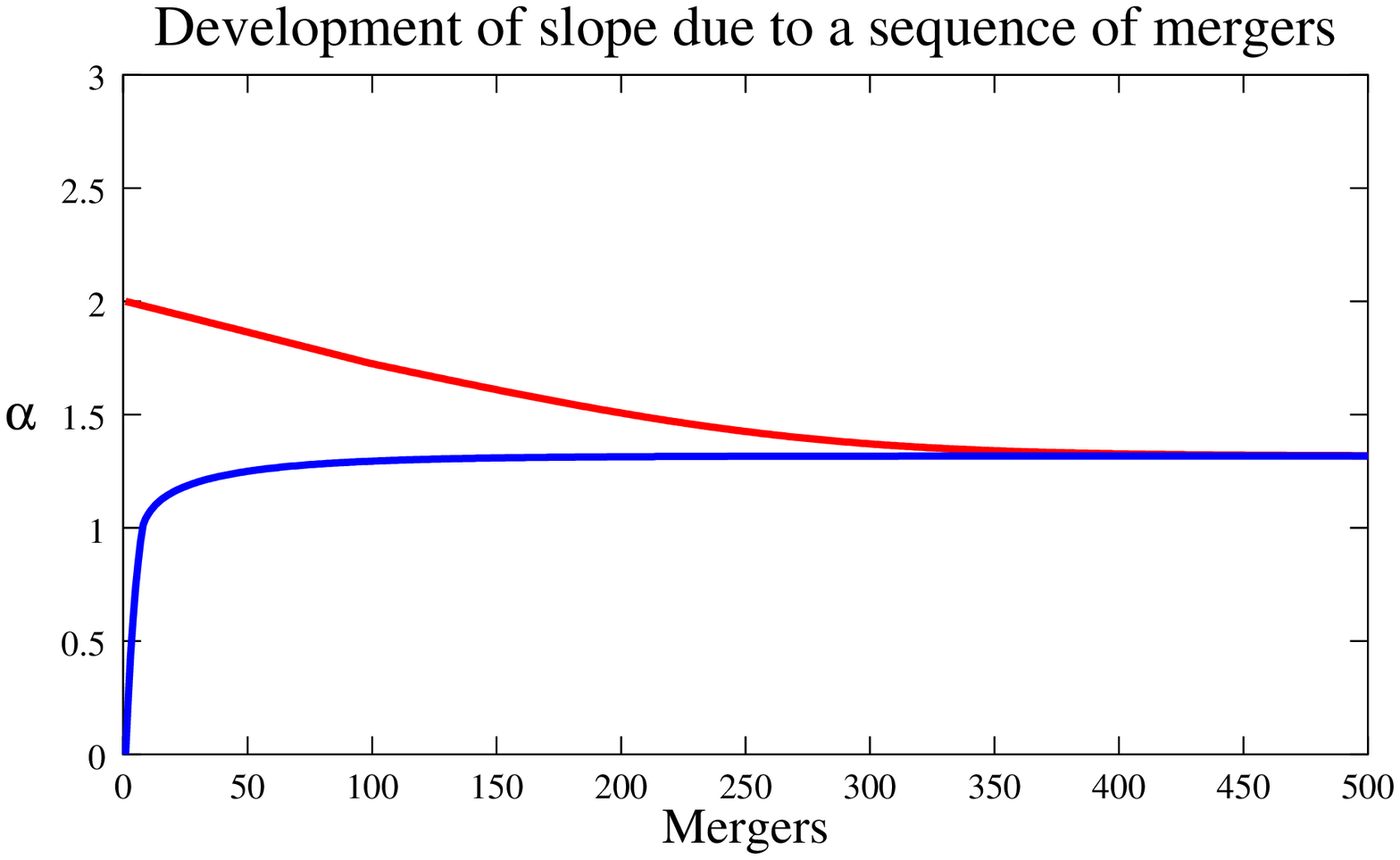}} 
{\includegraphics{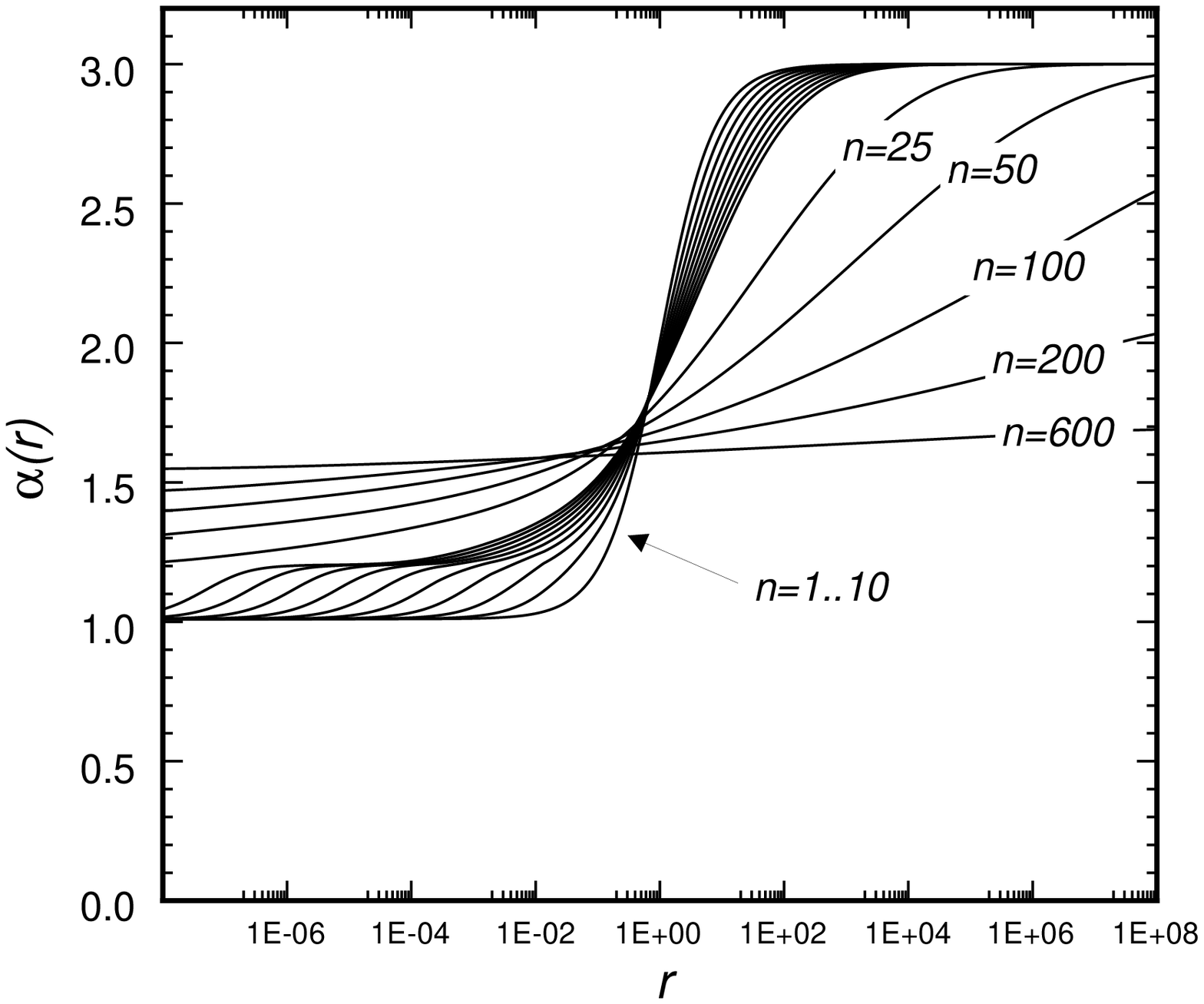}}
\caption{ 
{\it Left}:
Toy-simulation evolution of slope $\alpha$ at $r=0.1\rc$
due to a sequence of mergers $n=1,600$ with mass ratio $m/M=0.3$.
The initial profile is a generalized NFW with $\alpha$ either zero or 2.
When $\alpha<1$, the slope steepens rapidly to $\alpha>1$ within a few 
mergers 
(\S 2),
and then it converges slowly from either side towards an asymptotic value
(\S 4).
{\it Right}:
Corresponding evolution of slope profile $\alpha(r)$, starting
with $\alpha=1$ at $r \ll \rc$.
A power-law region develops below the radius where $\Delta\alpha=0$
(near $\rc$),
with a slope that grows slowly from unity to the asymptotic value.
	      } 
\label{fig:aas}
\end{figure}

\section{Discussion}
\label{sec:disc}

Our analysis demonstrates that the way to maintain
a flat core is by disrupting satellites outside the core. This may be
achieved if the cores of satellite halos are puffed up  
due to gas processes. 
As an example, the following speculative scenario is based on
enhanced feedback due to tidal compression. 
Consider a satellite made of DM and $\sim 10\%$ baryons passing
through the halo core towards a turn-around on the other side.
Assume that the baryons have already cooled and contracted into the
satellite center.  The satellite loses its outer DM layers in the outer halo
such that when it enters the halo core it is baryon-rich.
In the core, the tides compress the satellite, creating shocks and an
efficient burst of star formation.  Before turn-around on the other side,
there is time for the resulting supernovae to blow out the satellite
gas.  If the remaining satellite loses half its mass in this blow-out,
its density drops by a factor between 8 to infinity, depending on whether
the gas expulsion is adiabatic or impulsive.
Thus, the remaining satellite becomes much more susceptible to tidal
stripping, which disrupts it completely before it re-enters the halo core.

In our recent work, we address the different problems within the 
successful cosmological framework of CDM by appealing to inevitable 
feedback effects.  In Maller \& Dekel (2002) we address the 
angular-momentum catastrophe, where simulations including gas
produce disks significantly smaller than the galactic disks observed, and
with a different internal distribution of angular momentum.
We first construct a toy model for the angular-momentum buildup by mergers 
based on tidal stripping and dynamical friction, which helps us understand
the origin of the spin problem as a result of over-cooling in satellites.  
We then incorporate a simple model of feedback, motivated by Dekel \& Silk 
(1986), and find that it can remedy the discrepancies, and in particular
explain the low baryon fraction and angular-momentum profiles in dwarf 
disk galaxies.
Feedback effects may also provide the cure to the missing dwarf problem,
where the predicted number of dwarf halos in CDM is much larger than the 
observed number of dwarf galaxies (Bullock, Kravtsov \& Weinberg 2000).

Another approach (e.g., Hogan \& Dalcanton 2000)
is to appeal to a Warm Dark Matter (WDM) scenario,
despite the fact that it requires fine-tuning of the particle mass to
$\simeq 1~keV$.  The main feature of WDM is the partial suppression of 
small halos, which should help maintaining a core, but simulations of 
halos in WDM seem to still show inner cusps (Bullock, Kravtsov \& Colin 2001).  
While the explicit merger picture modeled
above may be invalid in this case, the gravitational processes involved
in the halo buildup still mimic a similar behavior.  
We note that the tidal compression in the core may amplify density
perturbations and make them behave like merging satellites.

The suppression of small halos in WDM may remedy the spin catastrophe
by avoiding over-cooling (Sommer-Larsen \& Dolgov 2001), 
but it is harder to see how it would explain 
the angular-momentum profile in galaxies. Furthermore,
while the number of dwarfs is already suppressed in WDM,
the addition of minimum feedback effects is likely to cause an overkill,
where the number of dwarfs is predicted to be
much lower than observed (J. Bullock, private comm.).

The success of our toy model in matching several independent observations
indicates that it indeed captures the relevant elements of the complex
processes involved, and in particular that feedback effects may indeed
provide the cure to all three problems of galaxy formation in CDM.



\end{document}